# ANALYSIS OF AN IN-SERVICE RUPTURE OF THE INNER FIRST WALL OF TORE SUPRA

R. Mitteau[1], P. Chappuis[1], G. Martin[1], S. Rosanvallon[2]

[1] Association Euratom-CEA, Département de Recherches sur la Fusion Contrôlée,
CEA/Cadarache, F-13108 SAINT PAUL LEZ DURANCE, France
[2] Département d'Etude des Réacteurs,
CEA/Cadarache, F-13108 SAINT PAUL LEZ DURANCE, France

**ABSTRACT**

The last experimental campaign of Tore Supra was also the last one with a set of plasma facing components mainly dating from the assembly of the machine in 1988. This campaign was interrupted a few days ahead of schedule because of the rupture of one coolant tube of the Inner First Wall (IFW). The rupture was observed by an abrupt increase of the pressure in the vessel. It was caused by a water leak which occurred two seconds after the end of the lower hybrid heating pulse during the current ramp - down. Investigations show that the accident is not related to the operation of the machine or a weaker element but rather to a localised heat flux exceeding the heat removal capability of the IFW.

## 1. INTRODUCTION : THE INNER FIRST WALL IN TORE SUPRA

The inner first wall of Tore Supra is a large area inner bumper, which is actively cooled by a pressurised water loop (2.8 MPa, temperature 150°C, velocity 3 m/s). The water flows in steel tubes (internal diameter 14 mm) with a square external section. Graphite or carbon fibre reinforced carbon (CFC) tiles are brazed on this heat sink through a 2 mm thick pure copper plate which absorbs the differential dilatation between the CFC and the steel. More details on this component can be found in former publications where it was largely described [1-4]. The IFW failed on november 17$^{th}$, 1999 after the end of the lower hybrid heating pulse during the current ramp - down, with a large water leak. The present paper focuses on the technological analysis of this accident, whereas plasma physic aspects are explored in [5], where photographs of the damaged component are also visible.

In vessel inspection shows that the failure occurred on one of the new panels installed in 1996. The tile is eroded a few tenth of millimetre, with the fibre structure of the CFC clearly visible. This overheating pattern is quite different from the hit pattern of runaway electrons in the equatorial plane and from heating leading edges that are observed at numerous locations. Neighbouring tiles are unharmed, apart from one just under which is only slightly marked. The copper plate has melted over one third of the surface of the braze bond, and a two millimetres gap remains where the copper once stood. Melted copper has leaked underneath. A reddish spot is observed around the location of the side, indicating that the failure may have occurred over more than one shot. The rest of tile is still strongly bonded to the heat sink, with no apparent change of its radial position. Another location is strongly metallised by braze material, indicating that it may have neared accident too (see section 2.). Those IFW panels underwent extensive non-destructive testing during manufacturing. The stainless steel wall thickness under that tile was measured to 1.3 mm (nominal = 1.5 mm), and the thermal transient tests with infrared monitoring (test-bed SATIR, [6]) were normal. The accident is therefore not related to a weak location.

Comparable erosion patterns were observed in 1997 on the IFW, although not associated to water leaks. A fully satisfactorily explanation could not be found at that time and operation proceeded without further alarms. Even before the installation of the new panels, the poloidal

**Corresponding author :**  Raphaël Mitteau - *Mailing adress, see above*
*phone* (+0/+33) 442 25 43 72      *fax* (+0/+33) 442 25 49 90      *Email* mitteau@drfc.cad.cea.fr

rows of first generation of IFW located around the equatorial plane suffered heavy damages. The interpretation was then hampered by the brittleness of the graphite tiles (which cracked when subjected to excessive thermal shocks) and the poor quality of many brazed bonds. These points show that the phenomenon is not new in Tore Supra, but that it only became critical with the increased use of lower hybrid additional heating associated with the lengthening of the discharges.

## 2. LOCATION OF THE FAILURE

The failure occurred 10° under the equatorial plane. The IFW is not perfectly toroidal, and the new panels were installed a few millimetres forward to be tested more thoroughly. The failure is located on the most advanced panel (see Fig. 1), and the second most advanced is also strongly damaged. It is stunning to see that tiles recessed from only 5 tenth of millimetres are untouched. This give hints of the narrowness of the phenomenon involved. The location of the accident indicates that it is bound to be related to the plasma, the failed tile being the one defining the last closed flux surface. In a poloidal section, the contact should be in the equatorial plane, but slight deformations of the plasma frontier may locate the contact point on the lower row of tiles that was actually touched

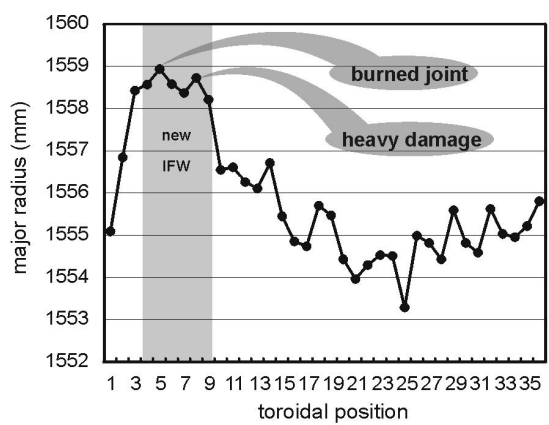

fig. 1 : toroidal profile of the Inner First Wall

## 3. THERMAL ANALYSIS

The lower hybrid additional heating was applied during 10 seconds. The time constant of the IFW is 10 seconds, which means that the stabilisation time of the structure is around 30 seconds. A heat flux considerably higher than the nominal value has therefore to be applied to damage the structure in the small time laps of 10s. Comparison of the two values gives information of that actual heat flux.

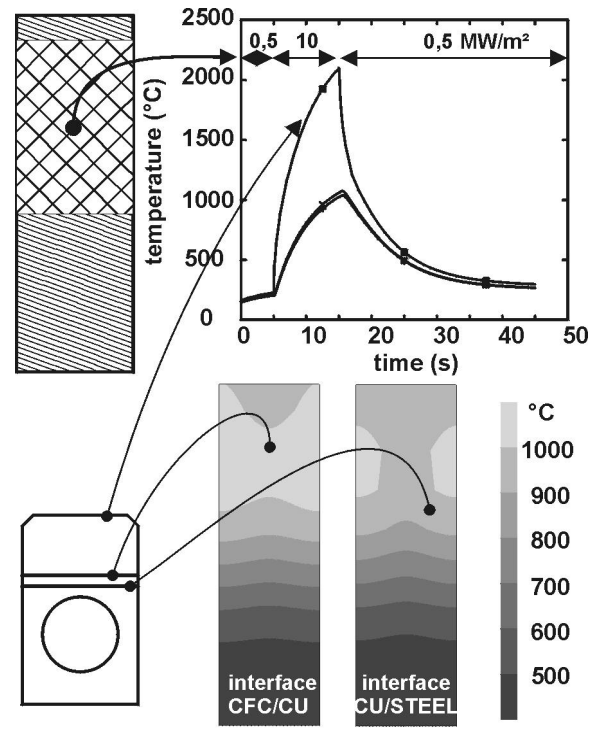

Fig. 2 : time - evolution of the temperature and temperature of brazed bonds at t = 15 s

A thermal transient calculation (Fig. 2) shows that this condition is obtained for a mean incident heat flux of 10 MW/m² over half the surface of the tile. This is ten times more than the nominal value of the structure (which was designed for 1 MW/m²) and five time more than the toughest test realised on it (2 MW/m² on the EB200 facility). Under these conditions, the interface between copper and the steel tube reaches the copper melting temperature of 1083°C in 10 seconds, necessary to explain the ejection of the copper. At the same time, the copper plate under the normal heat flux of 0.5 MW/m² is still under 500°C. This is the reason why the tile can have stayed well bonded to the heat sink, the interfaces still holding strongly in this area. With this combination of events, the flux incoming on the heavily loaded part of the tile is still passed on to the copper and to the stainless steel, increasing the power load on it and leading ultimately to the failure of the inner tube before complete detachment of the tile.

Two mechanisms are possible for the rupture of the stainless steel tube : firstly an overheating of the tube, leading to a cracking of the stainless steel, and secondly a critical heat flux in the water of the cooling circuit, which occur at 6 MW/m² incident for the hydraulic conditions of that shot. At that stage, it is not possible to distinguish between the two (see section 5 for further investigations about this.).

## 4. THEORETICAL HEAT FLUX DEPOSITION

A 10 MW/m² local heat flux is not normal on the IFW for 4 MW of additional power. The IFW is designed to fit closely the plasma shape and distribute the heat flux over a large surface. The heat flux pattern on the IFW was thoroughly investigated [7-9], both experimentally and by means of simulation. It was shown that a fraction of the power is

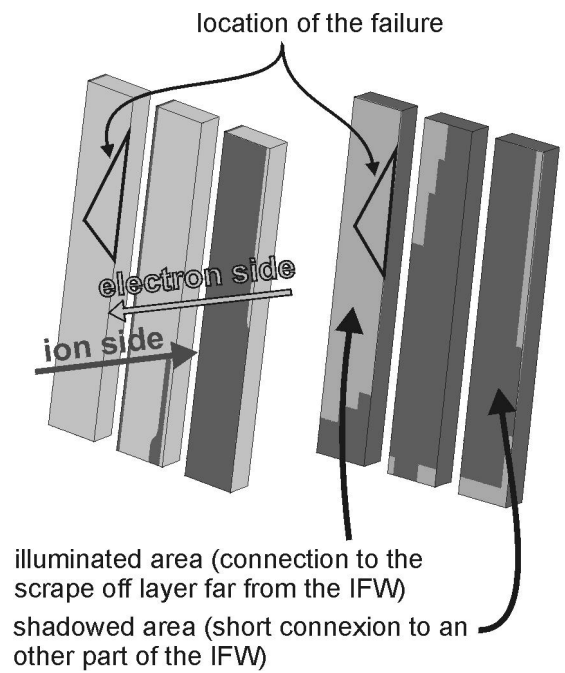

3.a : drift sides  3.b : IFW-shadowed areas
Fig. 3 : particle deposition on the IFW

deposited "abnormally", apparently as heat flux flowing along the field lines with a short e-folding lenght (a few millimetres as compared to the usual centimetres). A similar effect may have occurred here, in some exaggerated way. Sheath effects like surbrillance [10] may have played a role to concentrate even more power on the tile, but such mechanisms are unlikely to have been the trigger for the accident as they are not observed usually on this component.

The simulation gives some hints on the heat flux reaching the tile (Fig. 3). The tile is connected to the electron side, thus enabling the hypothesis of supra thermal electrons [5]. The eroded location is actually wetted by the heat flux, not being shadowed by the IFW itself as other tiles (Fig 3.b). A precise value of e-folding lenght can however hardly be given, because it would be a result of the adjustment of the simulation to the observation of the surface temperature field. As the parameter has an order of magnitude smaller than a millimetre, the correlation would require to know the geometry of the wall and the plasma frontier with an accuracy smaller than that, which is not the case.

## 5. POST-MORTEM METALLURGICAL EXAMINATION

The damaged structure was investigated by metallography. The copper has indeed been blown-up, possibly by water vapour coming with the 2.8 MPa pressure from the inside of the tube. A crater is observed where the copper once stood. The tile and the square tube have retained their external shape and are covered by a thin copper layer. Many cracks are found in the stainless steel tube, where the wall thickness is the smallest. Some of these cracks are emerging, which was the cause for the water leak. Non - emerging cracks show all stages of the formation. The cracks are initiated on the copper side and progress towards the inside, speaking for a rupture caused by excessive heat flux rather than by critical heat flux. The cracks are intergranular, reminiscent of a slow crack progression. This speaks also in favour of a progressive damaging of the structure, well compatible to the flux excess, whereas a critical heat flux is a faster event. Copper is present as dendrites in the cracks. Cracking may have be facilitated by the presence of liquid copper, which penetrates the grain boundaries of the stainless steel, causing their embrittlement.

non emerging cracks     emerging crack

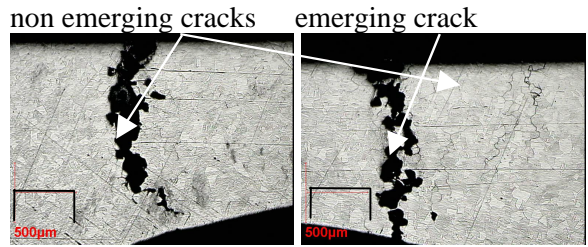

Fig. 4 : microstructure of the failed structure

A pattern of corrosion is observed in the tube around the crack, indicating the steel has heated up to around 600°C. There is however no loss of material and the corrosion is more likely to be a consequence of the accident than its cause. No changes in the structure of the carbon fibre reinforced carbon of the tile are observed.

## 6. CONCLUSION

The water leak of the IFW was caused by an excessive heat flux, between 5 and 10 time the nominal heat flux of the component. This heat flux is deposed abnormally within a very narrow depth of the scrape off layer, the investigations regarding this heat flux being reported elsewhere [5]. Rupture is attributed to an overheating of the stainless steel heat sink, which eventually cracked under the water pressure of the cooling water loop. The cracking of the steel was probably facilitated by an embrittlement caused by liquid copper. Physics issues related to this phenomenon will be addressed by Tore Supra with the new set of plasma facing components, CIEL, which are currently being installed. The local heat removal capacity is increased by a factor of 10, which is allowed by the use of a copper based heat sink. The surface temperature will be monitored by a set of infrared cameras, aiming at forbidding thermal excursions. These two elements give confidence that localised heat deposition will be better monitored and absorbed, which will facilitate the study of plasma scenario aimed at long discharges.

## 7. REFERENCES

[1] M. Lipa, P. Chappuis and P. Deschamps, Brazed graphite for actively cooled plasma facing components in Tore Supra, description, tests and performance, Fusion technology, Vol. 19, july 1991, pp. 2041-2048

[2] J.J. Cordier, P. Chappuis, D. Chatain et Al., Effect of misalignment and braze flaws on the Tore Supra inner first wall behaviour, fusion technology 1992, pp. 232 - 236

[3] M. Lipa, et Al., Development and fabrication of a new generation of CFC-brazed plasma facing component for Tore Supra, Report EUR/CEA-FC-1550, July 1995

[4] J. Schlosser et Al., In service experience feedback of the Tore Supra actively cooled inner first wall, Fusion Engineering and Design 27(1995)203-209.

[5] G. Martin, R. Mitteau, Y. Peysson, Fast Electron dynamic in the Tore Supra Plasma Edge, presented at 27th EPS 2000, to appear

[6] R. Mitteau, S. Berebi, P. Chappuis et Al., Non destructive testing of actively cooled plasma facing component by means of thermal transient excitation and infrared imaging, Fusion technology 1996, Vol 1, p.443-446, presented at the 19th SOFT, Lisbon, 16-20 September 1996.

[7] R. Mitteau, M. Chantant, P. Chappuis, et Al., Heat flux deposition pattern on the inner first wall of Tore Supra, Fusion technology 1998, Vol 1, p.129-132, presented at the 20th SOFT, Marseille, 7-11 September 1998.

[8] R. Mitteau, D. Guilhem, B. Riou et Al., Investigation of Power Deposition on Large Surfaces, Experiments and Simulations of the Tore Supra Inner First Wall, ECA Vol. 23J(1999)1001-1004, presented at the 26 EPS Conf. on Contr. Fusion and Plasma Physics, Maastricht, 14-18 june 1999.

[9] R. Mitteau, Ph. Chappuis, Ph. Ghendrih et Al., Self shadowing, gaps and leading edges on Tore Supra's inner first wall, J. of Nucl. Mat. NM, presented at 14th PSI, to appear

[10] Guilhem D., et Al, Observation and interpretation of thermal instabilities at the front face of actively cooled limiters in Tore Supra, 22nd European Conference on Controlled Fusion and Plasma Physics - Bournemouth, U.K., 3-7 July 1995